\documentclass{emulateapj}
\usepackage{float}

\begin{document}

\title{Quiescent Compact Galaxies at Intermediate Redshift in the COSMOS Field II. The Fundamental Plane of Massive Galaxies}

\author{H. Jabran Zahid$^1$\footnote{email: zahid@cfa.harvard.edu}, Ivana Damjanov$^2$, Margaret J. Geller$^1$ \& Igor Chilingarian$^{1,3}$}

\affil{$^1$Smithsonian Astrophysical Observatory, 60 Garden St., Cambridge, MA 02138, USA}
\affil{$^2$Harvard-Smithsonian Center for Astrophysics - 60 Garden Street, Cambridge, MA 02138}
\affil{$^3$Sternberg Astronomical Institute, Moscow State University, 13 Universitetsky prospect, 119992 Moscow, Russia}

\begin{abstract}

We examine the relation between surface brightness, velocity dispersion and size$-$the fundamental plane$-$for quiescent galaxies at intermediate redshifts in the COSMOS field. The COSMOS sample consists of $\sim150$ massive quiescent galaxies with an average velocity dispersion $\sigma \sim 250$ km s$^{-1}$ and redshifts between $0.2<z<0.8$. More than half of the galaxies in the sample are compact. The COSMOS galaxies exhibit a tight relation ($\sim0.1$ dex scatter) between surface brightness, velocity dispersion and size. At a fixed combination of velocity dispersion and size, the COSMOS galaxies are brighter than galaxies in the local universe. These surface brightness offsets are correlated with the rest-frame $g-z$ color and $D_n4000$ index; bluer galaxies and those with smaller $D_n4000$ indices have larger offsets. Stellar population synthesis models indicate that the massive COSMOS galaxies are younger and therefore brighter than similarly massive quiescent galaxies in the local universe. Passive evolution alone brings the massive compact quiescent COSMOS galaxies onto the local fundamental plane at $z = 0$. Therefore, evolution in size or velocity dispersion for massive compact quiescent galaxies since $z\sim1$ is constrained by the small scatter observed in the fundamental plane. We conclude that massive compact quiescent galaxies at $z\lesssim1$ are not a special class of objects but rather the tail of the mass and size distribution of the normal quiescent galaxy population.

\end{abstract}


\section{Introduction}

Observations of the structural properties of galaxies across cosmic time are critical for understanding how galaxies form and evolve. Most massive galaxies observed in the universe are not actively forming stars, i.e. they are quiescent. Quiescent galaxies exist at $z\sim4$ \citep{Fontana2009} and begin to dominate the massive galaxy population at $z = 2\sim3$ \citep{Ilbert2013, Muzzin2013}. The massive quiescent galaxies observed at $z\sim2$ have effective radii that are on average smaller than the local massive quiescent galaxy population \citep{Daddi2005, Trujillo2007, Zirm2007, Toft2007, Buitrago2008, vanDokkum2008b, Damjanov2011, vanderWel2014}. It is not clear whether the observed average size growth of quiescent galaxies since $z\sim2$ is due to the growth of individual galaxies \citep[e.g.,][]{Cimatti2008, Taylor2010} or to the addition of larger galaxies to the quiescent galaxy population at later times \citep[e.g.,][]{Valentinuzzi2010, Cassata2011, Cassata2013, Carollo2013}.

Several theoretical scenarios have been proposed to explain the origin and evolution of massive compact quiescent (MCQ) galaxies. Compact galaxies may form at $z\gtrsim2$ from gas-rich major mergers \citep{Khochfar2006} and/or instabilities in clumpy disks \citep{Elmegreen2008, Dekel2014}. At late times they may also form from tidal interactions \citep{Bekki2001, Chilingarian2009}. They can grow in size through minor mergers and accretion \citep{vanderWel2009, Naab2009,  Shih2011, Newman2012} and/or feedback driven adiabatic expansion \citep{Fan2008, Fan2010}. These growth mechanisms are evoked to account for both the larger average size of quiescent galaxies today as compared to $z\sim2$ and the putative dearth of MCQ galaxies in the local universe.

The number density of MCQ galaxies in the local universe probed by Sloan Digital Sky Survey (SDSS) is apparently 2-3 orders of magnitude smaller than at $z\sim1-2$ \citep{Trujillo2009, Taylor2010}. These estimates are consistent with an evolutionary scenario where MCQ galaxies form at high redshift and the number density rapidly declines as they grow in size since $z\sim1$ \citep[e.g.][]{vanDokkum2010, Newman2012}. However, number density estimates of MCQ galaxies in dense regions in the local universe \citep{Valentinuzzi2010, Poggianti2013} and some studies at intermediate redshifts (\citealt{Carollo2013}; \citealt{Damjanov2014}; Damjanov et al. 2015, submitted, hereafter Paper I) find no rapid decline. Instead, these studies show that the number density of MCQ galaxies remains roughly constant since $z\sim1$. {Thus, MCQ galaxies may not significantly grow in size since $z<1$ or they may be produced at a rate that compensates for their growth.} The relation between structural and kinematic properties of MCQ galaxies may provide important clues for understanding their origin and evolution. Here we examine the fundamental plane, i.e. the relation between surface brightness, central velocity dispersion and effective radius, for a sample of MCQ galaxies at $0.2 < z < 0.8$.

Galaxies in virial equilibrium show a relation between luminosity, size and velocity dispersion. \citet{Djorgovski1987} and \citet{Dressler1987} first demonstrated that quiescent galaxies in the local universe exhibit a tight correlation between surface brightness, effective radius and velocity dispersion$-$the fundamental plane (FP). The scatter in the FP is small \citep[$\sim0.05$ dex;][]{Bernardi2003, Saulder2013} and the FP appears to extend across all early-type galaxies in the local universe \citep{Misgeld2011}. 

The physical basis of the FP is the virial equilibrium of stellar systems dynamically supported by random motions. The observed properties of surface brightness, effective radius and velocity dispersion serve as proxies for the virial mass density, virial radius and virial velocity, respectively. As such, the FP is tilted relative to the expected virial relation \citep[e.g.,][]{Bernardi2003} and the FP evolves with redshift \citep{Treu2005, Holden2010, Saglia2010, Fernandez-Lorenzo2011, vandeSande2014}. Some of the evolution observed in the FP may possibly be attributed to evolution in the ratio between stellar and dynamical mass \citep{Beifiori2014, PeraltadeArriba2014}, though this is not the dominant effect.

We examine the fundamental plane for massive compact quiescent galaxies at $z<1$. In Section 2 we describe the data and methods and Section 3 contains the results. We discuss the results in Section 4 and we conclude in Section 5. We adopt the standard cosmology $(H_{0}, \Omega_{m}, \Omega_{\Lambda})$ = (70 km s$^{-1}$ Mpc$^{-1}$, 0.3, 0.7), AB magnitudes and a \citet{Chabrier2003} IMF.

\section{Data and Methods}

\subsection{Data Selection}

\begin{figure*}
\begin{center}
\includegraphics[width=2\columnwidth]{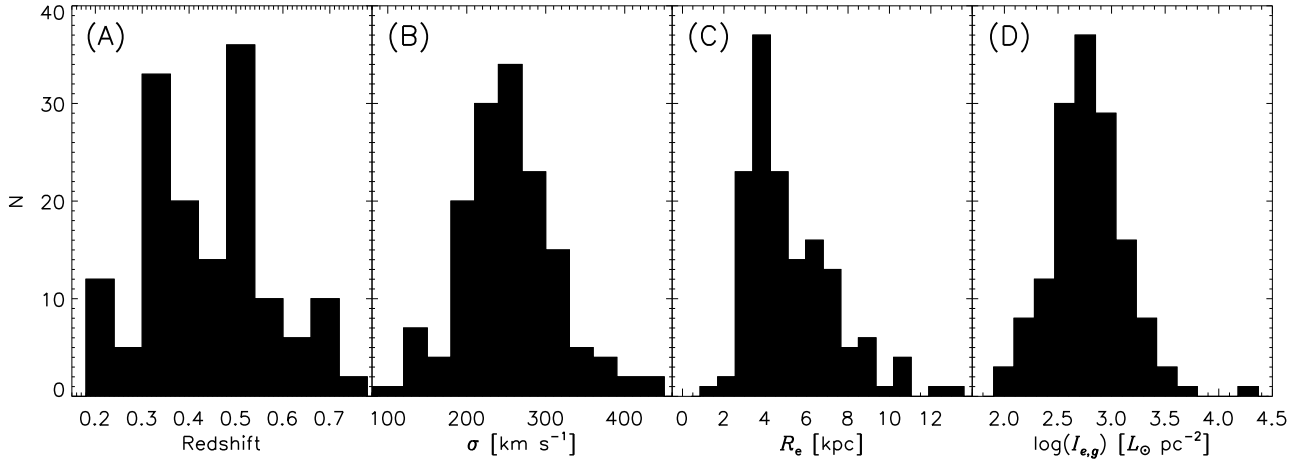}
\end{center}
\caption{(A) Redshift, (B) velocity dispersion, (C) circularized half-light radius and (D) surface brightness distribution of the 148 galaxies in the COSMOS sample.}
\label{fig:hist}
\end{figure*}

We examine galaxies in the 1.6 deg$^2$ COSMOS field \citep{Scoville2007}. All galaxies that we select in the COSMOS field have \emph{HST} ACS imaging \citep{Koekemoer2007} UV to IR multi-band photometry \citep[and references therein]{Ilbert2013} and SDSS{/BOSS spectroscopy to the SDSS and BOSS survey magnitude limits of $r<17.77$ and $i<19.9$, respectively \citep{Ahn2014}} We cross reference objects that have structural parameters given in the morphology catalog of \citet{Sargent2007}\footnote{http://irsa.ipac.caltech.edu/data/COSMOS/tables/morphology /cosmos\_morph\_zurich\_1.0.tbl}, multi-band photometry from the catalog of \citet{Ilbert2013}\footnote{http://vizier.cfa.harvard.edu/viz-bin/VizieR?-source=J/A+A /556/A55} and SDSS/BOSS spectroscopy by position matching objects within a 0.5 arcsecond radius. {We select galaxies in the intermediate redshift range $0.2 <z<0.8$ classified as quiescent by \citet{Ilbert2013} based on a rest-frame color-color selection ($NUV - r$ versus $r -J$). These selection criteria yield a sample of 161 galaxies.}

{We require robust measurements of velocity dispersions in order to derive the FP. From the cross-matched sample of 161 galaxies, we select galaxies with velocity dispersions $>70$ km s$^{-1}$ and velocity dispersion errors $<100$ km s$^{-1}$.} From examination of the spectra, we find that two galaxies in the sample have strong emission lines. Both galaxies exhibit emission line ratios consistent with active galactic nuclei \citep{Baldwin1981, Kauffmann2003, Kewley2006}. We remove these two galaxies from the sample. The final sample from which we derive the FP consists of 148 galaxies. We refer to this sample as the COSMOS sample.

{In Section 2.3 we investigate sample selection bias. The most restrictive selection is imposed by the SDSS/BOSS spectroscopic target selection. To assess the effects of this selection, we compare the COSMOS sample of 148 galaxies with a sample of quiescent galaxies in the COSMOS field with measured spectroscopic redshifts compiled from publicly available data \citep{Davies2015}. We restrict the comparison sample to galaxies with HST size measurements, stellar masses $>10^{10} M_\odot$ and redshifts in the range of $0.2<z<0.8$. This larger sample consists of 2970 galaxies. }

\subsection{Measured Properties}

The SDSS team derives redshifts and velocity dispersions according to \citet{Bolton2012}. Redshifts are determined from fitting template spectra at a range of trial redshifts. The central velocity dispersion is determined by comparing the observed spectra with model spectra which are redshifted to the galaxy redshift and convolved to the instrument resolution. Each model spectra is successively broadened to larger velocity dispersions in steps of 25 km s$^{-1}$. The best-fit velocity dispersion is determined by fitting for the velocity dispersion at the minimum chi-squared based on the measured chi-squared values at each 25 km s$^{-1}$ broadening step. 

We correct the velocity dispersion measured in the 2" (BOSS) and 3" (SDSS) apertures to the measured effective radius using the \citet[see their Equation 2]{Jorgensen1995} correction. The median correction applied to the sample is $\sim$0.04 dex. The typical error in the velocity dispersion for the COSMOS sample is $\sim30$ km s$^{-1}$. Figure \ref{fig:hist}A and \ref{fig:hist}B show the redshift and corrected velocity dispersion distribution, respectively. 

To constrain the stellar population age, we measure the $D_{n}4000$ index directly from the SDSS spectra. The $D_{n}4000$ index is an age sensitive spectral feature defined as the ratio of flux in two spectral windows adjacent to the $4000\mathrm{\AA}$ break \citep[for definition see][]{Balogh1999}. 

The effective radii of galaxies in the COSMOS sample are measured from \emph{HST} ACS imaging by \citet{Sargent2007}. The $\lesssim0.1"$ \emph{HST} resolution corresponds to a physical length of $\lesssim0.7$ kpc at $z\sim0.75$. \citet{Sargent2007} fit each surface brightness profile with a single \citet{Sersic1968} profile model using GIM2D \citep{Simard2002}. The formal uncertainty on size is typically $\lesssim 0.005$ dex. Because these errors are small, we ignore the uncertainty in size. We correct the measured semi-major half-light radius to the circularized averaged half-light (effective) radius given by
\begin{equation} 
R_e = a_{50}\sqrt{\frac{b}{a}}
\end{equation}
where $a_{50}$ is the semi-major half-light radius and $\frac{b}{a}$ is the semi minor-to-major axis ratio. \citet{vanderWel2014} find that galaxy sizes depend on the wavelength of observation. In order to compare with the local $g$-band FP, we correct the galaxy sizes to the rest-frame $g$-band effective wavelength \citep[$\lambda= 4686\AA$;][]{Stoughton2002}, using the correction given by \citet[see their Equation 2]{vanderWel2014}. The circularized and color corrected radius is typically $\sim0.04$ dex smaller on average than the measured semi-major effective radius. This correction does not change any of the major conclusions of this work. In Figure \ref{fig:hist}C we show the corrected effective radius distribution for the COSMOS sample.

We determine the surface brightness in the SDSS {rest-frame} $g, r, i, z-$bands by synthesizing photometry in these bands from the measured multi-band UV to IR SED. We determine k-corrected and reddening corrected magnitudes by fitting stellar population synthesis models of \citet{Bruzual2003} to the observed spectral energy distribution using the LePHARE\footnote{http://www.cfht.hawaii.edu/$\sim$arnouts/LEPHARE/lephare.html} code written by Arnout S. \& Ilbert O. \citep[for details see][]{Arnouts1999, Ilbert2006b}. 

The surface brightness in mag arcsec$^{-2}$ is 
\begin{equation}
\mu_{e} = m + 5 \, \mathrm{log}(r_e) + 2.5 \, \mathrm{log}(2\pi) - 10 \, \mathrm{log}(1+z),
\end{equation}
where $m$ is the reddening and k-corrected magnitude, $r_e$ is the radius measured in arcseconds and the final term is the cosmological surface brightness dimming correction. We convert the surface brightness to $I_e$, measured in $L_\odot$ pc$^{-2}$ using 
\begin{equation}
\mathrm{log} \left( \frac{I_e }{L_\odot ~ \mathrm{pc}^{-2}} \right) = 0.4\left(   M_\odot + 21.572 - \frac{\mu_e}{\mathrm{mag ~ arcsec}^{-2}} \right),
\label{eq:convert}
\end{equation}
where $M_\odot = 5.12$\footnote{http://classic.sdss.org/dr5/algorithms/sdssUBVRITransform.html} is the $g-$band solar absolute magnitude. The stellar mass in LePHARE represents the scale factor between the best-fit SED and the observed luminosity; the error estimate accounts for observational and some systematic uncertainties associated with the SED fitting procedure. The typical observational uncertainty on stellar masses is $\lesssim0.1$ dex, though the systematic errors may be larger \citep[e.g.][]{Conroy2010}. We adopt 0.1 dex as the error estimate on surface brightness. In Figure \ref{fig:hist}D we show the surface brightness distribution of the COSMOS sample. The results are independent of the photometric band. For ease of comparison with previous results, the analysis and results are based on the {rest-frame} $g-$band surface brightness. We give the measured and derived sample properties in Table \ref{tab:data}.

For the local benchmark, we compare with the orthogonal fit FP derived by \citet[][HB09 hereafter]{Hyde2009b} using $\sim50,000$ galaxies in the SDSS. We anticipate that some of the evolution in the FP is due to passive evolution of galaxies \citep[e.g.,][]{Treu2005} so we rewrite the HB09 FP in terms of surface brightness 
\begin{equation}
\mathrm{log}(I_{e,g}) = A \mathrm{log}(\sigma_e) + B \mathrm{log}(R_e) + C.
\label{eq:fit}
\end{equation}
Here, $\sigma_e$ and $R_e$ are measured in km s$^{-1}$ and kpc, respectively. We convert the FP parameters from HB09 into surface brightness units of $L_\odot$ pc$^{-2}$. The best-fit parameters for the local $g-$band FP are $A = 1.84$, $B = 1.31$ and $C=0.96$.

There is no single definition for compact galaxies. We identify galaxies as compact using the \citet{Barro2013} classification:
\begin{equation}
\mathrm{log}\left( \frac{M_\ast}{R_e^{1.5}} \right) \ge  10.3~M_\odot ~ \mathrm{kpc}^{-1.5}.
\label{eq:compact}
\end{equation}
Here, $M_\ast$ is the stellar mass in solar mass units. We adopt the stellar mass estimate given in the \citet{Ilbert2013} catalog. Using this definition, 57\% (85/148) of the galaxies in Table \ref{tab:data} are compact. 

\subsection{Sample Bias}
\label{sec:bias}

\begin{figure*}[t!]
\begin{center}
\includegraphics[width=2\columnwidth]{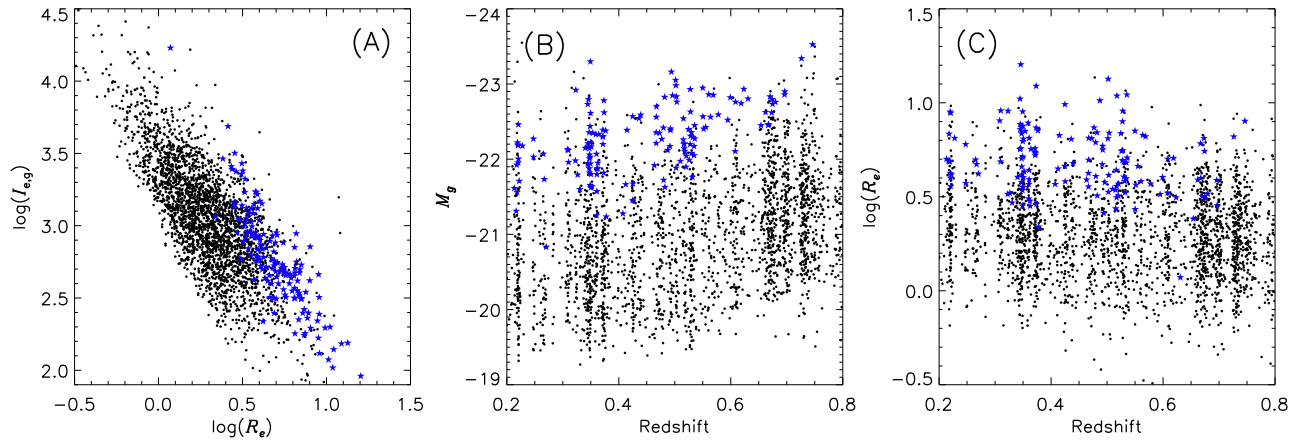}
\end{center}
\caption{(A) Surface brightness as a function of size, {(B) absolute $g$-band magnitude as a function of redshift and (C) size as a function of redshift. The blue stars and black points are the 148 galaxies in the COSMOS sample and the comparison sample of quiescent galaxies, respectively (see Section 2.1).}}
\label{fig:sb_dist}
\end{figure*}

Figure \ref{fig:sb_dist}A shows the surface brightness distribution as a function of size for the COSMOS sample (blue stars) relative to a similarly selected comparison sample (black points; see Section 2.1 for details of sample selection). The COSMOS sample is biased towards the largest galaxies and at a fixed size the selected sample populates the high surface brightness envelope of the full sample distribution. {Figures \ref{fig:sb_dist}B and \ref{fig:sb_dist}C show the absolute $g$-band magnitude and size as a function of redshift, respectively. The brightest and largest objects are selected across the redshift range.} The COSMOS sample is subject to the SDSS/BOSS target selection and the limits of SDSS/BOSS spectroscopy; the SDSS/BOSS target selection criteria lead to a selection bias towards high surface brightness galaxies. {Furthermore, as expected, Figure \ref{fig:sb_dist}B shows that increasingly brighter galaxies are selected with increasing redshift. Because of the SDSS/BOSS target selection, the galaxies in the COSMOS sample are drawn from the massive, high surface brightness tail of the quiescent galaxy distribution. The selection bias results in a large fraction of compact galaxies in the COSMOS sample.}

\section{Results}

\subsection{The Fundamental Plane}

\begin{figure}[t!]
\begin{center}
\includegraphics[width=\columnwidth]{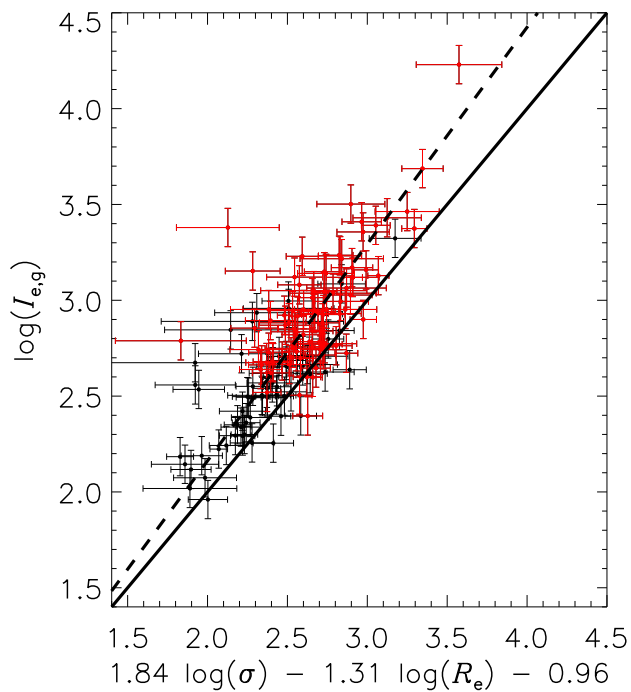}
\end{center}
\caption{Fundamental plane for the COSMOS sample. The solid line is the HB09 relation; the dashed line is the best fit FP for the COSMOS sample. The red points are compact galaxies.}
\label{fig:fpnocorr}
\end{figure}

Figure \ref{fig:fpnocorr} shows the relation between velocity dispersion, effective radius and surface brightness for the 148 galaxies in the COSMOS sample. The red points indicate compact galaxies (see Equation \ref{eq:compact}). The compact galaxies in the sample typically have higher surface brightness, higher velocity dispersion and smaller size. Thus, they populate the upper right hand part of the figure. The solid line is the local relation from HB09. 

The data show a clear offset from the local relation; for a given velocity dispersion and effective radius, galaxies in the COSMOS sample are brighter than local galaxies. The dashed line is a fit to the data (Equation \ref{eq:fit}). The fit is an orthogonal regression implemented in the \emph{sixlin.pro} IDL routine in the astronomy users library. The best-fit parameters are A = $2.09 \pm 0.12$, B = $-1.49 \pm 0.09$ and C = $-1.19 \pm 0.07$. The errors are based only on the dispersion of the data and do not account for observational uncertainties.


\begin{figure*}[ht!]
\begin{center}
\includegraphics[width=2\columnwidth]{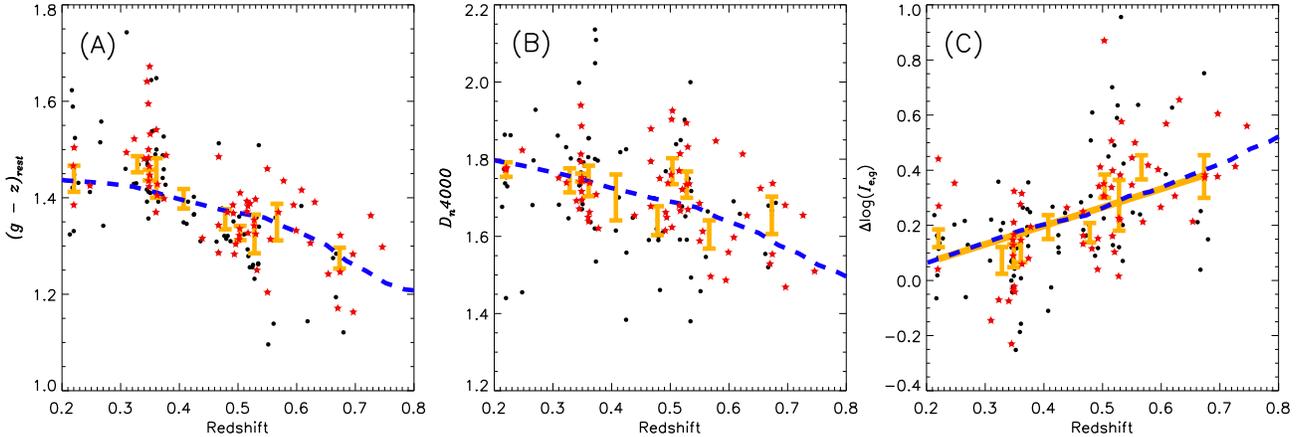}
\end{center}
\caption{(A) Rest-frame $g-z$ color, (B) $D_{n}4000$ index and (C) FP offsets for the COSMOS sample as a function redshift. {The black points and red stars are the individual galaxies in the COSMOS sample and the red stars denote compact galaxies.} The orange points are the median (A) rest-frame $g-z$ color, (B) $D_{n}4000$ index and (C) FP offsets in 10 equally populated bins of redshift. The errors bars are bootstrapped. The dashed blue curves show a model of a passively evolving galaxy where star formation began at $z\sim1.7$ and ceased at $z\sim1.3$. The solid orange line in (C) is a linear fit to the binned data.}
\label{fig:fpres}
\end{figure*}

\subsection{Quiescent Evolution}

Here we demonstrate that the surface brightness offsets of COSMOS galaxies relative to the local galaxy population (Figure \ref{fig:fpnocorr}) can be explained by simple quiescent evolution. We define the surface brightness offsets as the difference between the observed galaxy surface brightness of the COSMOS sample and the surface brightness calculated from the best-fit local FP:
\begin{equation}
\Delta \mathrm{log}(I_{e,g}) = \mathrm{log}(I_{e, m}) - \left[A \mathrm{log}(\sigma_{e, m}) + B \mathrm{log}(R_{e, m}) + C\right].
\label{eq:res}
\end{equation}
Here, $I_{e,m}$, $\sigma_{e,m}$ and $R_{e,m}$ are the measured surface brightness, velocity dispersion and effective radius for the COSMOS sample and $A$, $B$ and $C$ are best-fit local FP parameters taken from HB09 (see Section 2).

A rank correlation shows that the FP offsets, $\Delta \mathrm{log}(I_{e,g})$, are correlated with redshift ($6.8 \sigma$ significance), the rest-frame $g-z$ color\footnote{The offsets are strongly correlated with other rest-frame colors (i.e. $g-r$, $g-i$ and $r-i$) in the same sense; bluer galaxies show larger offsets.} ($6.5 \sigma$ significance), and $D_{n}4000$ index ($3.4\sigma$ significance). Bluer galaxies with smaller $D_n4000$ indices have larger surface brightness offsets. Both the rest-frame $g-z$ color and the $D_n4000$ index are also correlated with redshift; galaxies at higher redshifts are bluer and have smaller $D_n4000$ index. These trends are consistent with the simple interpretation that at a fixed combination of velocity dispersion and size, the higher redshift galaxies in the COSMOS sample are younger and therefore brighter and bluer than their local counterparts. {The SDSS/BOSS target selection partly contributes to these trends (see Section 2.3)} Thus mere quiescent evolution contributes to the observed offset between the COSMOS sample and the local FP. 

\begin{figure}
\begin{center}
\includegraphics[width=\columnwidth]{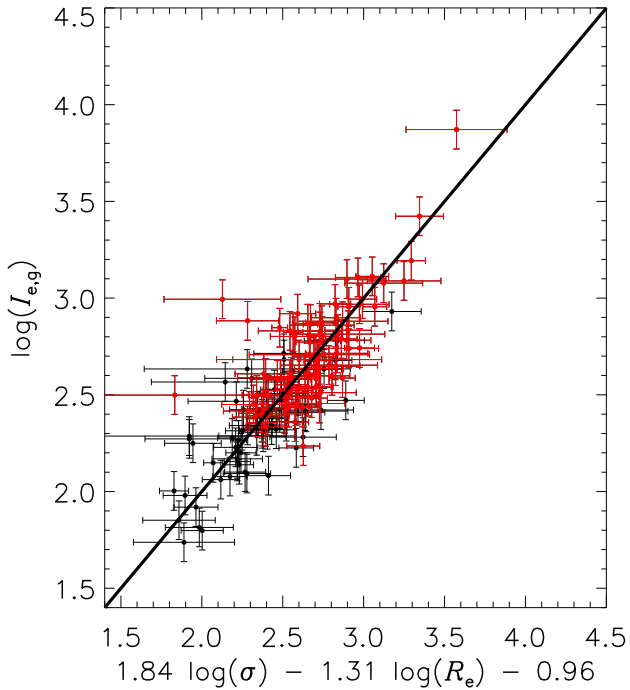}
\end{center}
\caption{Zero redshift fundamental plane accounting for quiescent evolution of the COSMOS sample. The solid line is the local fundamental plane relation from HB09. The red points are compact galaxies.}
\label{fig:fpcorr}
\end{figure}

The luminosity of a quiescent galaxy decreases as it passively evolves. To quantify this effect, we model a passively evolving galaxy using the Flexible Stellar Population Synthesis (FSPS) model \citep{Conroy2009a, Conroy2010}. FSPS generates both synthetic photometry and spectra as a function of time for an input star formation and metallicity history. We model both the $g-$band luminosity and the rest-frame $g-z$ color of a passively evolving galaxy based on synthetic magnitudes matched to the SDSS filters. We calculate the $D_{n}4000$ index directly from the synthetic spectra which have spectral resolution comparable to the SDSS. We model a galaxy with constant star formation rate of $\sim400$ $M_\odot$ yr$^{-1}$ for 1 Gyr at solar metallicity. The star formation rate is set to match the median stellar mass of the COSMOS sample. For a constant metallicity and constant and continuous star formation history, evolution of the $g-z$ color, $D_{n}4000$ index and the \emph{relative} evolution of the luminosity only depend on the duration of star formation. We tried star formation histories that spanned 0.5 - 2 Gyr in duration with similar results. The only free parameter in the model is the formation redshift.

We constrain the formation redshift of the model galaxy by fitting the median rest-frame $g-z$ color and $D_{n}4000$ index distribution of the COSMOS sample. Figures \ref{fig:fpres}A and \ref{fig:fpres}B show the rest-frame $g-z$ color and $D_{n}4000$ index as a function of redshift, respectively. In each panel the black points and red stars are the respective properties of individual galaxies in the COSMOS sample. The orange points are the median of the respective properties in 10 equally populated bins of redshift and the errors are bootstrapped. The dashed blue curves show the FSPS model. The rest-frame $g-z$ color and $D_{n}4000$ index distributions of the COSMOS sample are broadly consistent with a passively evolving galaxy which began star formation at $z\sim1.7$ and ceased star formation at $z\sim 1.3$. This redshift interval corresponds to 1 Gyr.

Figure \ref{fig:fpres}C shows the FP offsets, $\Delta \mathrm{log}(I_{e,g})$, as a function of redshift. The black points and red stars are FP offsets for individual galaxies in the COSMOS sample and the orange points are the median of the FP offsets in 10 equally populated bins of redshift. The errors are bootstrapped. The solid orange line is a fit to the median data:
\begin{equation}
\Delta \mathrm{log}(I_{e,g}) = (-0.071 \pm 0.049) + (0.67 \pm 0.12) z.
\label{eq:corr}
\end{equation}
The bootstrapped errors on the median are propagated through to the fit parameters. We refer to this relation as the luminosity evolution correction. The dashed blue curve is the $g-$band luminosity of the FSPS model galaxy normalized to the median $g-$band luminosity of the COSMOS sample $L_g \sim 10^{10.9} L_\odot$. The FP offsets plotted in Figure \ref{fig:fpres}C are relative to the local relation. The consistency between the change in luminosity of a passively evolving galaxy and the FP offsets means that passive luminosity evolution alone accounts for the offset between the FP we derive from the COSMOS data and the local FP from HB09 (dashed and solid line in Figure \ref{fig:fpnocorr}, respectively).

To demonstrate that passive evolution alone is sufficient to explain the offset of the COSMOS sample from the local FP, we apply the luminosity evolution correction given by Equation \ref{eq:corr} to the data. We emphasize that this evolution correction is consistent with the luminosity evolution of a passively evolving galaxy which ceased star-formation at $z \sim 1.3$ (see Figure \ref{fig:fpres}C). Figure \ref{fig:fpcorr} shows the FP for the COSMOS sample brought to $z=0$ by accounting for the quiescent luminosity evolution shown in Figure \ref{fig:fpres}. The scatter of the data in Figure \ref{fig:fpcorr} is $\sim0.15$ dex. Based on the observational uncertainties in velocity dispersion ($\sim0.07$ dex) and luminosity ($\sim 0.1$ dex) the intrinsic scatter is $\lesssim 0.1$ dex. The luminosity corrected data are consistent with the local FP. Passive evolution of galaxies in the COSMOS sample is sufficient to place them on the local FP by $z\sim0$.

\section{Discussion}

We examine the FP for a sample of galaxies spanning the redshift range of $0.2 < z < 0.8$. Due to the strong selection bias towards high surface brightness objects, MCQ galaxies comprise a large fraction of the COSMOS sample we examine. Passive evolution alone brings the COSMOS sample onto the local FP by $z=0$. {The physical basis of the fundamental plane is the virial equilibrium of quiescent galaxies. Thus we conclude that MCQ galaxies are virialized systems.}

{Several studies examining the FP at intermediate and high redshifts have recognized the importance of evolution of stellar populations. \citet{Treu2005} report that offsets from the fundamental plane for galaxies in the redshift range of $0.2<z<1.2$ are anti-correlated with stellar mass; lower mass galaxies have larger offsets. They attribute these trends to downsizing. Lower mass galaxies are younger and therefore have smaller mass-to-light (M/L) ratios. To account for evolution of stellar populations and varying M/L ratios, \citet{Bezanson2013} derive the ``mass" FP substituting stellar mass surface density for surface brightness. In contrast to the observed offsets in the FP, they find very small offset between the mass FP at $z\sim2$ and the local relation. \citet{vandeSande2014} quantify the evolution of the zero-point of the FP out to $z\sim2$. They find that $\Delta \mathrm{log} (M/L_{g}) \propto (-0.49 \pm 0.03)z$.} 

{We conclude that the FP offsets are due to the evolution of stellar populations. In particular, we find that passive evolution alone can bring galaxies in the COSMOS sample onto the local FP relation. Thus, $\Delta \mathrm{log}(I_{e,g}) = - \Delta \mathrm{log} (M/L_g)$. Based on comparisons of our data with stellar population synthesis models, we determine that $\Delta \mathrm{log}(I_{e,g}) \propto (0.69 \pm 0.12)z$ (see Equation \ref{eq:corr}). The evolution we measure is consistent (1.5$\sigma$) with evolution reported by \citet[also see references therein]{vandeSande2014}. We note however that both studies are biased towards bluer objects and therefore the reported evolution may not be representative of the quiescent galaxy population. In general, our results are qualitatively consistent with previous studies examining the evolution of the FP. The novel aspect of this work is that it demonstrates that MCQ galaxies follow the same evolutionary trends as the general quiescent galaxy population.}

{Due to the SDSS/BOSS target selection, COSMOS galaxies examined in this study are outliers in the relation between stellar mass and size. However, the stellar population and kinematics of MCQ galaxies in the COSMOS sample are consistent with the local quiescent galaxy population. Thus, we conclude that MCQ galaxies at $0.2<z<0.8$ represent the extreme of the mass and size distribution of normal quiescent galaxies (see Figure \ref{fig:sb_dist}) and are not a unique class of objects. \citet{Saulder2015} reach similar conclusions based on a larger sample culled from the literature.}

Recent cosmological hydrodynamical simulations suggest that compact galaxies at $z\sim2$ do not form from unique physical mechanisms but rather are subject to the same formation processes as other galaxies \citep{Wellons2014}. Thus, both the origin and evolution of MCQ galaxies appear to be consistent with the conclusion that compact galaxies are the tail of the normal galaxy distribution.

If MCQ galaxies at $z<1$ passively evolve with little or no size growth, their descendants should be identifiable among the local galaxy population. However, the number density evolution of compact galaxies since $z\sim1$ remains unsettled. Several studies based on the SDSS claim that the number density of compact galaxies drops dramatically in the local universe \citep[e.g.][]{Shen2003, Trujillo2009, Taylor2010, Cassata2013}. This has lead to the suggestion that they grow significantly since $z\sim1-2$. If this is the case, the small scatter in the FP at $z<1$ means that any growth mechanism ostensibly moves galaxies along, not off, the FP. {Additionally, if galaxies do grow in size, the results of this study provide empirical constraints for theoretical analysis of galaxy growth \citep[see for example][]{Hopkins2010}.} However, it is possible that compact galaxies today are preferentially found in environments that are incompletely sampled by the SDSS \citep[see][]{Taylor2010}. Thus their number densities may be systematically underestimated in the local universe.

\begin{figure}
\begin{center}
\includegraphics[width=\columnwidth]{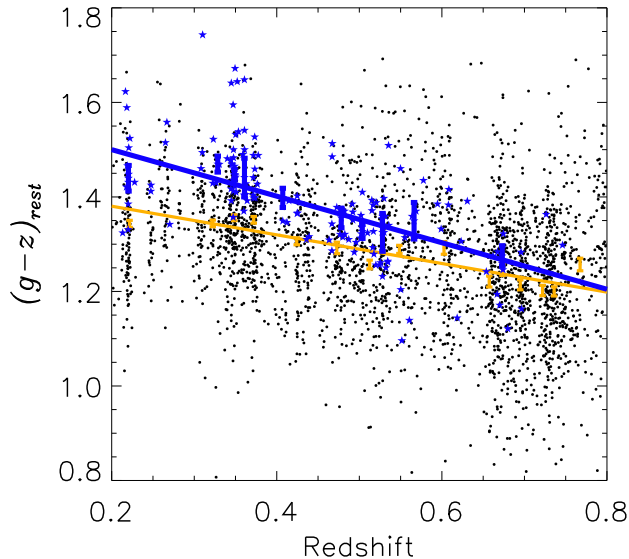}
\end{center}
\caption{Rest-frame $g-z$ color as a function of redshift for the COSMOS sample (blue stars) and the comparison sample (black points). The blue and orange points are the median rest-frame $g-z$ color in equally populated bins of redshift and the blue and orange lines are fits to the median color for the COSMOS sample and the comparison sample, respectively. The error bars are bootstrapped.}
\label{fig:color}
\end{figure}

In contrast to studies reporting a rapid decline in the number density of compact galaxies at $z<1$, several studies find that the number density remains roughly constant (e.g., \citealt{Valentinuzzi2010}; \citealt{Poggianti2013}; \citealt{Carollo2013}; \citealt{Damjanov2014}). The results of our companion paper (Paper I) are consistent with this conclusion. In this case, either galaxies do not grow in size or compact galaxies are produced at a rate that compensates for their growth. 

\citet{Carollo2013} conclude that evolution in the mass-size relation is dominated by newly formed quiescent galaxies, i.e. progenitor bias. They argue that a scenario where individual galaxies grow and new compact galaxies form at $z<1$ is inconsistent with the fact that compact galaxies are systematically redder at lower redshifts suggesting that the population is aging over time and that the normal quiescent galaxy population is bluer than compact galaxies indicating that the regular population has more recently shutdown star formation as compared to the compact population. 

Figure \ref{fig:color} compares the rest-frame $g-z$ color of the selected COSMOS sample, dominated by MCQ galaxies, with the comparison COSMOS sample (see Section 2.1 for sample description). The data are broadly consistent with the interpretation of \citet{Carollo2013}. MCQ galaxies are redder at late times and redder than the normal quiescent galaxy population. However, we note that the scatter in the rest-frame $g-z$ color and $D_n4000$ index is large; there are galaxies in the selected COSMOS sample that have colors and $D_n4000$ indices consistent with younger stellar populations. A combination of progenitor bias \emph{and} individual galaxy growth may be the basis for the evolution of the mass-size relation. Moreover, given the strong selection bias in the selected COSMOS sample, we probe only the extreme end of the mass distribution. Spectroscopically complete samples at $z<1$ will provide important constraints for the role of progenitor bias and individual galaxy growth in the evolution of the mass-size relation.


Comparison of the COSMOS sample with a model of a passively evolving galaxy shows that, on average, the data are consistent with a galaxy which formed stars from $z\sim1.7$ to $z\sim1.3$. However, we emphasize the galaxies in the COSMOS sample span a broad range of ages and star formation histories contributing to the large scatter in the rest-frame $g-z$ color and $D_n4000$ index (Figure \ref{fig:fpres}A and \ref{fig:fpres}B, respectively). The best-fit $z=1.7$ formation redshift is the average formation redshift of the sample.

Figure \ref{fig:fpcorr} shows that once we apply the evolutionary correction to luminosity given by Equation \ref{eq:corr}, the FP we derive from the COSMOS sample is consistent with the local relation. However, we have applied only an average correction for evolution which does not account for stellar population variations amongst galaxies at any particular redshift and the scatter in Figure \ref{fig:fpres} highlights the fact that quiescent galaxies are a heterogenous population. The selection bias of our sample complicates the analysis of any residual offsets between our evolved sample and the local FP (Figure \ref{fig:fpcorr}). A complete spectroscopic sample combined with sophisticated modeling of individual galaxies will allow for more robust and detailed kinematic studies.

\section{Summary and Conclusions}

We examine the relation between surface brightness, velocity dispersion and size$-$the fundamental plane$-$for a sample of massive compact quiescent galaxies in the COSMOS field spanning the redshift range of $0.2<z<0.8$.  Based on analysis of this COSMOS sample, we show that:

\begin{itemize}

\item{Massive compact galaxies at $z\lesssim1$ populate a tight fundamental plane relation similar to the general population of quiescent galaxies in the local universe. {This reflects the fact that massive compact galaxies are in virial equilibrium.}}

\item{We compare the COSMOS sample to a model of a passively evolving galaxy. The average properties of the COSMOS sample are consistent with a galaxy which started star formation at $z\sim1.7$ and ceased star formation at $z\sim1.3$. Accounting for passive evolution of the surface brightness brings the COSMOS sample onto the $z=0$ fundamental plane.}


\item{The data suggest that massive compact quiescent galaxies at $z<1$ are not a special class of objects; they are the high mass, high surface brightness tail of the normal quiescent galaxy population.}

\end{itemize}

In Paper I we show that the number density of compact galaxies remains constant for $z<1$. This study (Paper II) concludes that the compact galaxy population is the tail of the normal galaxy population. These studies taken together demonstrate the potential of combining abundance, structural and kinematic analyses for investigating the origin and evolution of the quiescent galaxy population.

\acknowledgements

We thank the anonymous referee for their careful reading of the manuscript. HJZ gratefully acknowledges the generous support of the Clay Postdoctoral Fellowship. ID is supported by the Harvard College Observatory Menzel Fellowship and the Natural Sciences and Engineering Research Council of Canada Postdoctoral Fellowship (NSERC PDF-421224-2012). IC acknowledges support from the Russian Science Foundation project \#14-22-00041. This research has made use of NASA's Astrophysics Data System Bibliographic Services.

Funding for SDSS-III has been provided by the Alfred P. Sloan Foundation, the Participating Institutions, the National Science Foundation, and the U.S. Department of Energy Office of Science. The SDSS-III web site is http://www.sdss3.org/.

SDSS-III is managed by the Astrophysical Research Consortium for the Participating Institutions of the SDSS-III Collaboration including the University of Arizona, the Brazilian Participation Group, Brookhaven National Laboratory, University of Cambridge, Carnegie Mellon University, University of Florida, the French Participation Group, the German Participation Group, Harvard University, the Instituto de Astrofisica de Canarias, the Michigan State/Notre Dame/JINA Participation Group, Johns Hopkins University, Lawrence Berkeley National Laboratory, Max Planck Institute for Astrophysics, Max Planck Institute for Extraterrestrial Physics, New Mexico State University, New York University, Ohio State University, Pennsylvania State University, University of Portsmouth, Princeton University, the Spanish Participation Group, University of Tokyo, University of Utah, Vanderbilt University, University of Virginia, University of Washington, and Yale University.

Based on observations made with the NASA/ESA Hubble Space Telescope, obtained at the Space Telescope Science Institute, which is operated by the Association of Universities for Research in Astronomy, Inc., under NASA contract NAS 5-26555.

\bibliographystyle{apj}
\bibliography{/Users/jabran/Documents/latex/metallicity}

\LongTables
\begin{deluxetable*}{ccccccc}
\tablewidth{450pt}
\tablecaption{Sample Properties}
\tablehead{\colhead{Redshift} &\colhead{$\sigma$} & \colhead{$R_e$} & \colhead{$\mathrm{log}(I_{e,g})$}  & \colhead{$\mathrm{log}(M_\ast)$} & \colhead{$(g-z)_{rest}$} & \colhead{$D_{n}4000$}  \\
 & [km s$^{-1}$] & [kpc]  & [$L_\odot$ pc$^{-2}$]  & [$M_\odot$]}
\startdata
0.218 & 199 & 6.83 & 2.35 & 11.25 & 1.42 & 1.86 \\ 
0.310 & 170 & 9.09 & 2.12 & 11.45 & 1.74 & 1.68 \\ 
0.346 & 290 & 16.00 & 1.96 & 11.68 & 1.47 & 1.73 \\ 
0.219 & 305 & 4.90 & 2.76 & 11.41 & 1.47 & 1.77 \\ 
0.349 & 254 & 6.24 & 2.55 & 11.40 & 1.44 & 1.67 \\ 
0.349 & 290 & 2.83 & 3.16 & 11.25 & 1.47 & 1.72 \\ 
0.345 & 359 & 7.22 & 2.40 & 11.65 & 1.64 & 1.67 \\ 
0.376 & 221 & 7.11 & 2.50 & 11.45 & 1.43 & 1.80 \\ 
0.228 & 233 & 6.48 & 2.50 & 11.42 & 1.43 & 1.86 \\ 
0.359 & 360 & 7.82 & 2.40 & 11.51 & 1.49 & 1.75 \\ 
0.220 & 196 & 3.99 & 2.92 & 11.30 & 1.39 & 1.78 \\ 
0.222 & 223 & 4.15 & 2.70 & 11.20 & 1.52 & 1.73 \\ 
0.349 & 269 & 7.53 & 2.54 & 11.53 & 1.44 & 1.75 \\ 
0.347 & 304 & 3.78 & 2.95 & 11.35 & 1.50 & 1.94 \\ 
0.347 & 252 & 6.86 & 2.52 & 11.67 & 1.59 & 1.74 \\ 
0.364 & 230 & 5.67 & 2.67 & 11.46 & 1.39 & 1.64 \\ 
0.696 & 300 & 3.87 & 3.22 & 11.45 & 1.28 & 1.68 \\ 
0.363 & 319 & 2.82 & 3.13 & 11.21 & 1.43 & 1.74 \\ 
0.308 & 233 & 8.03 & 2.29 & 11.39 & 1.46 & 1.86 \\ 
0.415 & 222 & 6.71 & 2.55 & 11.51 & 1.39 & 1.82 \\ 
0.578 & 254 & 3.71 & 3.15 & 11.60 & 1.44 & 1.85 \\ 
0.555 & 195 & 4.60 & 2.89 & 11.46 & 1.31 & 1.61 \\ 
0.436 & 251 & 6.41 & 2.65 & 11.50 & 1.31 & 1.65 \\ 
0.322 & 211 & 3.27 & 3.01 & 11.10 & 1.43 & 1.83 \\ 
0.348 & 249 & 4.57 & 2.66 & 11.14 & 1.41 & 1.77 \\ 
0.515 & 306 & 2.74 & 3.39 & 11.50 & 1.37 & 1.68 \\ 
0.680 & 405 & 3.28 & 3.32 & 11.05 & 1.12 & 1.69 \\ 
0.697 & 259 & 2.84 & 3.50 & 11.50 & 1.16 & 1.47 \\ 
0.530 & 261 & 4.85 & 2.82 & 11.45 & 1.40 & 1.83 \\ 
0.525 & 148 & 6.87 & 2.53 & 11.39 & 1.26 & 1.90 \\ 
0.670 & 107 & 3.17 & 3.38 & 11.24 & 1.17 & 1.53 \\ 
0.527 & 310 & 4.94 & 2.74 & 11.40 & 1.34 & 1.89 \\ 
0.218 & 141 & 2.75 & 2.75 & 10.70 & 1.26 & 1.36 \\ 
0.361 & 277 & 4.38 & 2.71 & 11.30 & 1.50 & 1.85 \\ 
0.667 & 316 & 6.07 & 2.66 & 11.10 & 1.19 & 1.52 \\ 
0.520 & 173 & 3.21 & 3.00 & 11.05 & 1.26 & 1.68 \\ 
0.623 & 368 & 3.13 & 3.43 & 11.55 & 1.31 & 1.81 \\ 
0.561 & 218 & 3.82 & 3.23 & 11.19 & 1.14 & 1.55 \\ 
0.528 & 288 & 7.95 & 2.62 & 11.70 & 1.35 & 1.73 \\ 
0.361 & 235 & 3.91 & 2.78 & 11.30 & 1.54 & 1.64 \\ 
0.360 & 234 & 3.69 & 2.94 & 11.30 & 1.37 & 1.65 \\ 
0.352 & 344 & 4.30 & 2.64 & 11.24 & 1.64 & 1.79 \\ 
0.220 & 267 & 4.21 & 2.78 & 11.30 & 1.43 & 1.75 \\ 
0.528 & 269 & 9.05 & 2.39 & 11.40 & 1.23 & 1.59 \\ 
0.425 & 236 & 3.82 & 2.76 & 11.20 & 1.37 & 1.56 \\ 
0.348 & 297 & 5.05 & 2.65 & 11.45 & 1.48 & 1.79 \\ 
0.265 & 225 & 4.98 & 2.68 & 11.35 & 1.51 & 1.68 \\ 
0.425 & 253 & 9.81 & 2.29 & 11.55 & 1.40 & 1.83 \\ 
0.531 & 84 & 3.83 & 2.79 & 11.21 & 1.36 & 1.75 \\ 
0.746 & 255 & 7.99 & 2.85 & 11.75 & 1.30 & 1.51 \\ 
0.350 & 295 & 4.01 & 2.76 & 11.38 & 1.53 & 1.74 \\ 
0.528 & 227 & 5.58 & 2.68 & 11.45 & 1.36 & 1.75 \\ 
0.662 & 328 & 3.89 & 3.12 & 11.25 & 1.27 & 1.55 \\ 
0.526 & 123 & 5.54 & 2.56 & 11.16 & 1.26 & 1.65 \\ 
0.467 & 265 & 4.22 & 2.94 & 11.44 & 1.29 & 1.79 \\ 
0.467 & 292 & 6.49 & 2.71 & 11.55 & 1.51 & 1.79 \\ 
0.631 & 322 & 1.18 & 4.23 & 11.60 & 1.39 & 1.65 \\ 
0.532 & 186 & 3.33 & 3.12 & 11.33 & 1.25 & 1.74 \\ 
0.516 & 138 & 4.39 & 2.85 & 11.26 & 1.29 & 1.59 \\ 
0.362 & 218 & 3.01 & 3.05 & 11.20 & 1.41 & 1.66 \\ 
0.502 & 299 & 6.46 & 2.85 & 11.67 & 1.31 & 1.76 \\ 
0.569 & 301 & 6.62 & 2.75 & 11.60 & 1.37 & 1.50 \\ 
0.595 & 262 & 3.26 & 3.23 & 11.41 & 1.39 & 1.49 \\ 
0.530 & 199 & 4.04 & 2.87 & 11.29 & 1.33 & 1.67 \\ 
0.535 & 186 & 11.03 & 2.14 & 11.45 & 1.51 & 1.49 \\ 
0.536 & 191 & 5.11 & 2.74 & 11.40 & 1.34 & 1.72 \\ 
0.378 & 207 & 2.19 & 3.06 & 11.05 & 1.49 & 1.62 \\ 
0.482 & 130 & 3.20 & 2.89 & 10.93 & 1.32 & 1.46 \\ 
0.412 & 259 & 4.46 & 2.62 & 11.20 & 1.35 & 1.70 \\ 
0.487 & 207 & 10.28 & 2.07 & 11.20 & 1.31 & 1.61 \\ 
0.393 & 254 & 1.52 & 3.28 & 10.67 & 1.39 & 1.83 \\ 
0.492 & 269 & 5.89 & 2.66 & 11.54 & 1.37 & 1.66 \\ 
0.270 & 144 & 4.19 & 2.34 & 10.57 & 1.34 & 1.93 \\ 
0.477 & 219 & 7.37 & 2.42 & 11.43 & 1.32 & 1.62 \\ 
0.492 & 288 & 5.63 & 2.71 & 11.37 & 1.35 & 1.77 \\ 
0.550 & 272 & 2.71 & 3.41 & 11.12 & 1.20 & 1.59 \\ 
0.351 & 241 & 6.42 & 2.50 & 11.15 & 1.41 & 1.68 \\ 
0.667 & 286 & 3.97 & 3.04 & 11.27 & 1.28 & 1.68 \\ 
0.502 & 243 & 13.38 & 2.19 & 11.65 & 1.36 & 1.77 \\ 
0.517 & 283 & 3.60 & 2.99 & 11.32 & 1.38 & 1.81 \\ 
0.517 & 228 & 3.14 & 3.14 & 11.38 & 1.33 & 1.67 \\ 
0.518 & 191 & 10.90 & 2.02 & 11.25 & 1.28 & 1.85 \\ 
0.267 & 275 & 4.71 & 2.60 & 11.32 & 1.56 & 1.80 \\ 
0.654 & 358 & 2.42 & 3.46 & 11.24 & 1.24 & 1.72 \\ 
0.463 & 210 & 3.89 & 2.84 & 11.24 & 1.31 & 1.59 \\ 
0.248 & 207 & 3.73 & 2.92 & 11.33 & 1.42 & 1.82 \\ 
0.619 & 187 & 5.08 & 2.94 & 11.05 & 1.14 & 1.66 \\ 
0.349 & 309 & 3.57 & 3.12 & 11.60 & 1.42 & 1.89 \\ 
0.673 & 138 & 6.48 & 2.67 & 11.26 & 1.26 & 1.67 \\ 
0.361 & 254 & 6.49 & 2.25 & 11.30 & 1.65 & 1.85 \\ 
0.608 & 204 & 5.03 & 2.95 & 11.62 & 1.42 & 1.60 \\ 
0.345 & 286 & 10.50 & 2.30 & 11.60 & 1.42 & 1.80 \\ 
0.310 & 366 & 4.86 & 2.72 & 11.48 & 1.49 & 1.75 \\ 
0.372 & 242 & 5.27 & 2.69 & 11.25 & 1.49 & 1.80 \\ 
0.220 & 209 & 8.95 & 2.22 & 10.95 & 1.33 & 1.44 \\ 
0.354 & 258 & 7.06 & 2.49 & 11.55 & 1.54 & 1.78 \\ 
0.349 & 292 & 8.98 & 2.66 & 11.85 & 1.45 & 1.69 \\ 
0.503 & 156 & 4.12 & 3.15 & 11.62 & 1.34 & 1.62 \\ 
0.340 & 365 & 4.02 & 2.90 & 11.39 & 1.48 & 1.69 \\ 
0.323 & 395 & 9.04 & 2.51 & 11.76 & 1.52 & 1.74 \\ 
0.221 & 251 & 8.83 & 2.34 & 11.50 & 1.45 & 1.79 \\ 
0.373 & 193 & 12.28 & 2.18 & 11.61 & 1.46 & 2.11 \\ 
0.372 & 269 & 5.46 & 2.73 & 11.45 & 1.44 & 2.14 \\ 
0.213 & 174 & 5.02 & 2.49 & 11.05 & 1.32 & 1.68 \\ 
0.326 & 238 & 3.31 & 2.93 & 11.05 & 1.43 & 1.65 \\ 
0.467 & 309 & 4.97 & 2.88 & 11.55 & 1.48 & 1.88 \\ 
0.494 & 424 & 2.59 & 3.69 & 11.75 & 1.28 & 1.65 \\ 
0.473 & 251 & 6.46 & 2.57 & 11.40 & 1.38 & 1.70 \\ 
0.726 & 299 & 6.61 & 2.95 & 11.82 & 1.36 & 1.65 \\ 
0.467 & 304 & 3.58 & 3.03 & 11.35 & 1.34 & 1.68 \\ 
0.608 & 216 & 3.32 & 3.05 & 11.16 & 1.38 & 1.69 \\ 
0.515 & 198 & 3.44 & 3.08 & 11.50 & 1.39 & 1.77 \\ 
0.439 & 280 & 4.76 & 2.93 & 11.59 & 1.32 & 1.65 \\ 
0.564 & 250 & 6.73 & 2.62 & 11.35 & 1.36 & 1.66 \\ 
0.221 & 230 & 3.59 & 2.94 & 11.40 & 1.50 & 1.78 \\ 
0.425 & 182 & 4.60 & 2.50 & 11.05 & 1.36 & 1.38 \\ 
0.374 & 431 & 2.90 & 3.37 & 11.47 & 1.40 & 1.71 \\ 
0.599 & 334 & 4.02 & 3.17 & 11.60 & 1.33 & 1.56 \\ 
0.347 & 237 & 7.45 & 2.26 & 11.04 & 1.36 & 1.75 \\ 
0.350 & 254 & 3.55 & 2.80 & 11.10 & 1.44 & 1.76 \\ 
0.349 & 270 & 3.42 & 2.84 & 11.08 & 1.46 & 1.76 \\ 
0.371 & 257 & 5.72 & 2.66 & 11.40 & 1.40 & 1.62 \\ 
0.345 & 232 & 7.69 & 2.36 & 11.37 & 1.39 & 1.73 \\ 
0.480 & 222 & 6.15 & 2.64 & 11.50 & 1.37 & 1.59 \\ 
0.360 & 216 & 3.09 & 2.77 & 11.07 & 1.40 & 1.81 \\ 
0.552 & 288 & 3.66 & 3.09 & 11.03 & 1.10 & 1.46 \\ 
0.674 & 256 & 5.53 & 2.88 & 11.50 & 1.25 & 1.55 \\ 
0.480 & 142 & 4.06 & 2.72 & 10.90 & 1.32 & 1.60 \\ 
0.482 & 330 & 6.94 & 2.70 & 11.70 & 1.37 & 1.75 \\ 
0.496 & 289 & 3.77 & 3.24 & 11.39 & 1.30 & 1.74 \\ 
0.373 & 192 & 7.37 & 2.24 & 11.25 & 1.53 & 1.53 \\ 
0.550 & 261 & 7.08 & 2.73 & 11.65 & 1.46 & 1.69 \\ 
0.510 & 183 & 3.47 & 2.96 & 11.07 & 1.32 & 1.84 \\ 
0.346 & 238 & 5.72 & 2.51 & 11.35 & 1.47 & 1.74 \\ 
0.491 & 302 & 4.71 & 2.77 & 11.40 & 1.38 & 1.62 \\ 
0.407 & 255 & 3.69 & 2.63 & 11.00 & 1.35 & 1.61 \\ 
0.534 & 272 & 4.80 & 2.71 & 11.15 & 1.26 & 1.38 \\ 
0.344 & 265 & 5.66 & 2.52 & 11.26 & 1.48 & 2.00 \\ 
0.247 & 236 & 6.02 & 2.60 & 11.30 & 1.41 & 1.46 \\ 
0.218 & 226 & 4.87 & 2.50 & 11.10 & 1.59 & 1.74 \\ 
0.504 & 252 & 6.91 & 2.68 & 11.60 & 1.41 & 1.93 \\ 
0.349 & 281 & 5.17 & 2.60 & 11.50 & 1.67 & 1.75 \\ 
0.502 & 201 & 3.35 & 2.95 & 11.18 & 1.32 & 1.90 \\ 
0.674 & 304 & 3.11 & 3.36 & 11.60 & 1.32 & 1.74 \\ 
0.330 & 219 & 2.94 & 2.95 & 11.06 & 1.47 & 1.80 \\ 
0.348 & 268 & 4.14 & 2.84 & 11.44 & 1.43 & 1.81 \\ 
0.217 & 220 & 4.89 & 2.40 & 11.13 & 1.62 & 1.77 \\ 
0.534 & 264 & 3.95 & 2.93 & 11.25 & 1.26 & 2.00 \\ 
0.372 & 233 & 5.38 & 2.71 & 11.20 & 1.40 & 2.05 \\ 
0.521 & 286 & 3.70 & 2.93 & 11.40 & 1.39 & 1.72 
\enddata
\label{tab:data}
\tablecomments{An electronic version of these data is available from HJZ upon request.}
\end{deluxetable*}

 \end{document}